%% file: main.tex
\documentclass[twocolumn,tighten]{aastex63}

\usepackage{times,natbib,graphicx,amsmath,multirow,xspace}
\usepackage{xcolor}
\usepackage{lineno}
\usepackage{booktabs}
\usepackage{placeins}
\usepackage{tabularx}

\newcommand{\nustar}{NuSTAR\xspace}

\newcommand{\rg}{$R_{g}$\xspace}
\newcommand{\risco}{$R_{\mathrm{ISCO}}$\xspace}

\newcommand{\relxillns}{{\sc relxillNS}\xspace}

\newcommand{\source}{\mbox{Sco X-1}\xspace}
\newcommand{\commentout}[1]{}

\shorttitle{Sco X-1}
\shortauthors{Li et al.}

\begin{document}

\title{NuSTAR’s Intentional Stray Light Observation of Scorpious X-1}

\author[0009-0005-8520-0144]{S.~Li}
\affiliation{Department of Physics \& Astronomy, Wayne State University, 666 West Hancock Street, Detroit, MI 48201, USA}
\affiliation{Independent Researcher}

\author[0000-0002-8961-939X]{R.~M.~Ludlam}
\affiliation{Department of Physics \& Astronomy, Wayne State University, 666 West Hancock Street, Detroit, MI 48201, USA}

\author[0000-0003-0440-7978]{M.~Sudha}
\affiliation{Department of Physics \& Astronomy, Wayne State University, 666 West Hancock Street, Detroit, MI 48201, USA}

\author[0000-0002-4024-6967]{M.~C.~Brumback}
\affiliation{Department of Physics, Middlebury College, Middlebury, VT 05753, USA}

\author[0000-0002-5341-6929]{D.~J.~K.~Buisson}
\affiliation{Independent Researcher}

\author[0000-0003-0870-6465]{B.~M.~Coughenour}
\affiliation{Department of Physics, Utah Valley University, 800 W. University Parkway, MS 179, Orem, UT 84508, USA}

\author[0000-0003-0331-3259]{A.~Di Marco}
\affiliation{INAF Istituto di Astrofisica e Planetologia Spaziali, Via del Fosso del Cavaliere 100, 00133 Roma, Italy}
\affiliation{Dipartimento di Fisica, Universit\`{a} degli Studi di Roma ``Tor Vergata'', Via della Ricerca Scientifica 1, I-00133 Roma, Italy}

\author[0000-0002-1984-2932]{B.~W.~Grefenstette}
\affiliation{Cahill Center for Astronomy and Astrophysics, California Institute of Technology, Pasadena, CA 91125, USA}

\author[0000-0001-8916-4156]{F.~La Monaca}
\affiliation{INAF Istituto di Astrofisica e Planetologia Spaziali, Via del Fosso del Cavaliere 100, 00133 Roma, Italy}
\affiliation{Dipartimento di Fisica, Universit\`{a} degli Studi di Roma ``Tor Vergata'', Via della Ricerca Scientifica 1, I-00133 Roma, Italy}

\author[0000-0003-4216-7936]{G.~Mastroserio}
\affiliation{Scuola Universitaria Superiore IUSS Pavia, Palazzo del Broletto, piazza della Vittoria 15, I-27100 Pavia, Italy}

\author[0000-0001-6304-1035]{S.~Rossland}
\affiliation{Center for National Security Initiatives, University of Colorado Boulder, Boulder, CO 80309, USA}

\begin{abstract}

We present the first spectral analysis of  Scorpius X-1 (Sco X-1) using intentional stray light (SL) observations taken by NuSTAR. Unlike focused observations that have high telemetry load when observing bright sources, intentional SL observations can help reduce the telemetry and reduce the effect of dead time, thereby maximizing the on-source exposure time; all of which are critical for extremely bright sources that exhibit short timescale variability like Sco X-1. The intentional SL  observation of Sco X-1, taken in 2023, captured the source primarily in the flaring branch (FB) of the Z track. We performed spectral modeling of the continuum and reprocessed emission. A combination of  thermal and Comptonization components (modeled with {\sc thcomp}) provided a robust fit to the continuum. We test both scenarios for Comptonized emitting regions arising from the accretion disk and close to the neutron star, which provides comparable fit statistics. Reflection was modeled with the \relxillns model, enabling measurements of disk inclination consistent with prior radio and IXPE studies and comparison of inner disk radius to the emission radii of the thermal components. Overall, the results from the intentional SL data provide comparable results to literature on the focused NuSTAR data of Sco X-1 in the FB or taken contemporaneously. The success of this observation demonstrates the capability of SL data to yield high-quality spectral constraints comparable to focused observations, offering a powerful avenue for studying bright X-ray binaries with NuSTAR.

\end{abstract}


\section{Introduction} \label{sec:intro}
Low-mass X-ray binaries (LMXBs) are a type of binary system that consists of a compact object and a donor star. The compact object can be a neutron star (NS) or black hole (BH) and the donor star has a mass $\lesssim 1\ M_{\odot}$. The compact object accretes matter from the donor star through Roche-lobe overflow and forms an accretion disk. These systems are some of the brightest X-ray sources in the sky. NS LMXBs can be classified into two categories: atoll and Z sources. These two classifications can be distinguished by the shape of their color-color diagram or hardness intensity diagram (HID), where atoll sources have an island-like shape with two states: `island' and `banana' state. Z sources have a Z shape and have three different branches: horizontal branch (HB), normal branch (NB), and flaring branch (FB). The vertices between HB, NB, and FB are called hard apex (HA) and soft apex (SA). 

The continuum emission of NS LMXBs can be modeled with a single temperature blackbody component that describes the NS surface and/or boundary layer (BL) with a multi-temperature blackbody component that describes the accretion disk \citep{Mitsuda_1984}. In addition, the soft photons emitted from the NS surface and/or inner region of the disk can Compton scatter off the high energy electrons in a central compact Comptonizing medium such as the BL or hot corona around the disk. In 1980s, an `Eastern' model \citep{Mitsuda_1989} and a `Western' model \citep{White_1988} were proposed to account for the Comptonization in LMXBs. The Eastern model assumes that the Comptonization arises from the BL and uses a multi-temperature disk component with a Comptonized single-temperature blackbody component to describe the spectra. Alternatively, the Western model assumes that the Comptonization arises from the disk and uses a single-temperature blackbody component with a Comptonization model on the mutli-temperature blackbody component to describe the spectra. However, both Eastern and Western models have difficulties describing the spectra in the soft state of NS LMXBs, so a hybrid model \citep{Lin_2007} that uses a power-law to describe the weak Comptonization tail in a thermally dominated spectral state was proposed. Therefore, NS LMXBs are commonly modeled by a combination of a single temperature blackbody component, a multi-temperature blackbody component, and a component for Comptonization \citep{DAi_2010,church_2006,Pandel_2008,Coughenour_2018,wang_2019}.

In LMXBs, when X-rays illuminate the disk, they often interact with the material in the innermost region of the disk by scattering, absorption, and re-emission that results in a reprocessed component in the observed spectrum known as ``reflection" \citep{Ross_2005}. The concept of X-ray reflection as a probe of the innermost accretion flow was first developed for active galactic nuclei (AGNs:\citealt{Fabian_1989,George_1991}), and was later extended to  Galactic X-ray binaries \citep{Miller_2002,Bhattacharyya_2007,Cackett_2008}. One of the most prominent feature for reflection is a broadened iron (Fe) K$\alpha$ line between 6.4 -- 6.97 keV, which has an intrinsically narrow emission line profile that is shaped by Doppler, special and general relativistic effects from the innermost disk \citep{Fabian_1989,Laor_1991, Fabian_2000}. Some other indications of reflection arise from lower energy emission features including oxygen Lyman alpha \citep{Madej_2011,Madej_2014} and Fe L-shell transitions \citep{Renee_2018} as well as the Compton hump at higher energies. From modeling the reflection spectrum, information about the disk, including the ionization state, electron density, chemical composition, inner disk radius, disk inclination, etc., can be obtained. See \cite{Ludlam24} for a brief review of the field of relativistic reflection in NS LMXBs and references therein.

Scorpius X-1 (Sco X-1) is the brightest persistent X-ray source in the sky and it was the first extra-solar X-ray source discovered \citep{Giacconi_1962}. The source is located at a distance of $2.13_{-0.26}^{+0.21}$ kpc away based on the Gaia DR2 parallax measurements \citep{Arnason_2021}, which is closer than the previous parallax measurements obtained with Very Long Baseline Array ($2.8\pm0.3$~kpc: \citealt{Bardshaw_1999}). The source is viewed at an inclination of $44^{\circ}\pm 6^{\circ}$ based on radio lobe motion \citep{Fomalont_2001}. Sco X-1 is a NS LMXB that is classified as a Z source. It is the prototypical ``Sco-like" Z source, which means that the source primarily occupies NB and FB, but rarely is found in HB \citep{Church_2012}.

Spectral studies of Sco X-1 have been performed using data from different X-ray telescopes including RXTE, HXMT, XMM-Newton, NuSTAR, INTEGRAL, and IXPE with diverse models \citep[e.g.,][]{White_1985,DAmico_2001,Bradshaw_2003,DA_2007,De_2009,Church_2012,Maiolino_2013,Titarchuk_2014,Reig_2016,Mazzola_2021,Alexander_2023,Ding_2023,Fabio_2024,Gouse_2025}. Observations from NuSTAR have shown a soft component attributed to a disk blackbody with $kT_{in} \sim 0.4$--$0.6$ keV and a blackbody from the neutron star surface at $kT_{bb} \sim 0.8$--$1.5$ keV \citep{Mazzola_2021,Fabio_2024}. A Comptonized component, modeled with {\sc nthcomp}, has an optical depth of 6--10 and a coronal electron temperature $\sim 3$ keV \citep{Mazzola_2021}. Additionally, a hard X-ray tail from above 30 keV and extended up to 200--300 keV without high energy cut off, which suggests that there is no hot corona present \citep{Ding_2023,Revnivtsev_2014}. The hard X-ray tail can be found in all branches throughout the Z track but becomes harder and weak from the HB, through NB, to the FB \citep{Revnivtsev_2014,Ding_2023} and the amplitude varies by more than a factor of 10 \citep{Revnivtsev_2014}. Exosat data revealed a broad iron line centered at 6.75~keV with an equivalent width that decreased from 50 eV to 25 eV during the FB \citep{White_1985}. The IXPE observation on 2023 August 23 has shown highly significant X-ray polarization for Sco X-1 with a polarization degree (PD) of $\sim 1.08\%$ and a polarization angle (PA) of $\sim 8^{\circ}\pm6^{\circ}$. The PA is misaligned with the radio jet axis by $\sim 46^{\circ}$, suggesting that the emission geometry is not jet dominated \citep{Fabio_2024}. In addition, the misalignment may also suggest relativistic precession or changes in coronal geometry across different accretion states. The observed polarization is consistent with expectation of a Comptonization component with an optical depth $\tau \sim 7$ and an electron temperature  $kT_e\sim 3$~keV. The PD is consistent with theoretical predictions for an optically thick, electron-scattering atmosphere viewed at an inclination of $\sim 44^{\circ}$ \citep{Fabio_2024}. 

\nustar \citep{Harrison_2013} is the first X-ray telescope that focuses hard X-rays from 3--79 keV using multilayer-coated mirrors. It provides a high spatial and energy resolution that fills the critical gap in hard X-ray astronomy above 10 keV with an energy resolution of 400~eV at 10 keV. \nustar has an open geometry that connects the optics to detectors that allows light from a roughly $1-4^\circ$ off axis angle to illuminate the detectors without being focused by the optics \citep{madsen_2017}. This is referred to as ``stray light" (SL). Despite the fact that SL observations have a lower signal to noise ratio (S/N) in comparison to focused data, it has unique advantages. Since SL observations usually occur serendipitously during a focused observation, there may be more SL data than focused data available for a specific target (e.g., see figure 1 in \citealt{songwei_2025} for long-term light curve of GX 340+0). In addition to SL observations potentially providing more data than focused observations, they may also span a longer time period—from NuSTAR’s launch in 2012 to the present—whereas focused observations are limited to specific scheduled times secured through Guest Observing cycles. Another advantage of SL is that the energy range is not limited by \nustar's mirrors since the photons do not pass through the optics. This means that spectra can be extended beyond 79 keV, which makes it a great way for observing hard X-ray sources \citep{Mastroserio_2022}. 

In some cases, SL observations are performed intentionally where observations of a source are scheduled off-axis to bypass the mirrors while maximizing the illumination of the detectors. This is important for the calibration of the effective area of \nustar, provides more accurate high-energy and high off-axis angle corrections, and better agreement between its focal plane modules (FPMs) across detectors \citep{madsen_2017}. In addition, intentional SL observation are often done for bright sources to reduce the telemetry load and avoid data loss; enabling longer on-source exposure time and to capture multiple orbital periods \citep{Grefenstette_2021}.

Given that \source is the brightest, persistent X-ray source in the sky, the high X-ray flux leads to extreme deadtime effects (due to pixels turning off to read out event triggers) during a  focused \nustar observation resulting in effective exposure times that are just 5\% of the time spent observing the target. Thus, observing \source as a stray light target could provide high-quality spectra while alleviating limitations to observing bright sources with \nustar. 
In this paper, we present the first spectral analysis of \nustar intentional SL data of \source. The observation was taken in August 2023 and covered the FB of the Z track. In \S2, we describe the data and background reduction of the source. In \S3, we discuss the spectral analysis including both continuum and reflection models. In \S4, we discuss the results and compare it with the literature prior to concluding.

\begin{figure}[t]
  \centering
  \includegraphics[width=0.4\textwidth, trim=5 0 0 0, clip]{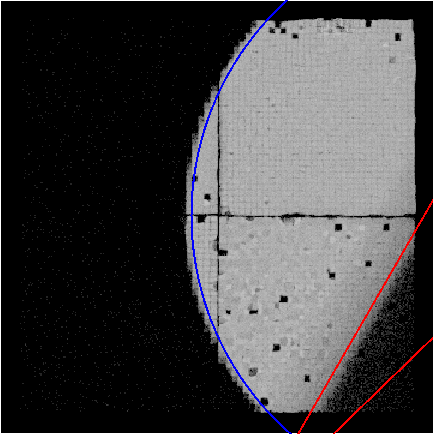}  
  \includegraphics[width=0.4\textwidth, trim=5 0 0 0, clip]{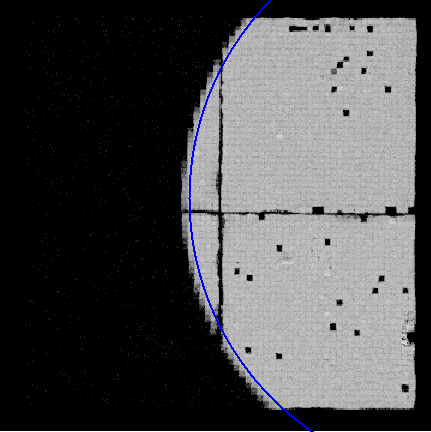}  
  \caption{ Detector images for \nustar FPMA (top) and FPMB (bottom) SL observations of \source. A circular source region (blue) was used for data extraction on both FPMs. An exclusion box region (red) was used to remove \nustar's shadow on FPMA. Region files are available on the StrayCats website.}
  \label{fig:detimage}
\end{figure}

\begin{figure*}[t!]
  \centering
  \includegraphics[width=\textwidth]{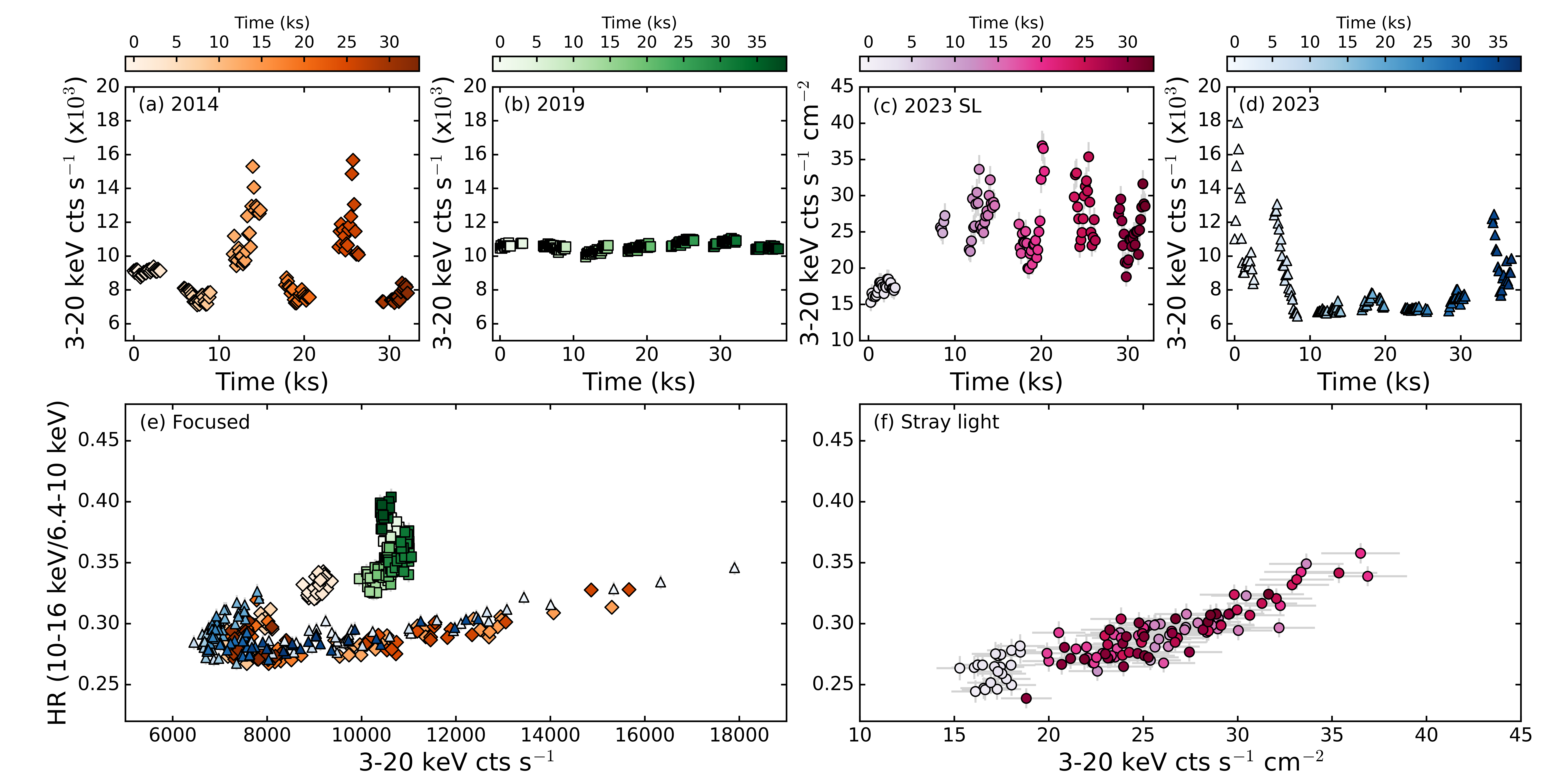}
  \caption{Top row shows the light curves of the NuSTAR focused (panels a, b, \& d) and SL (panel c) observations of Sco X-1 in chronological order. The color bar denotes the elapsed time since the start of each respective observation. Note that the SL observation in 2023 occurred directly prior to the focused observation. The bottom row shows the hardness-intensity diagrams of the (panel e) focused and (panel f) SL data. Only one FPM from each observation is shown for clarity. }
  \label{fig:HID}
\end{figure*}

\section{Data reduction} \label{sec:Data Reduction}
\nustar has observed \source on three occasions as a focused target (ObsIDs: 30001040002, 30502010002, 30902036002) and once as intentional SL (ObsID 10902336002). The \nustar SL observation occurred in 2023 directly prior to the focused observation presented in \cite{Fabio_2024}. The SL data were reduced using {\sc nustardas} v2.1.2 and \nustar CALDB v20231205. Since the SL observations were intentional, source photons illuminate both FPMs though the illuminating area differs between FPMA and FPMB (6.75 cm$^{2}$ and 7.73 cm$^{2}$, respectively) due to the orientation of \nustar. We provide the detector images for the intentional SL observation of \source with the region files denoted in Figure~\ref{fig:detimage}.

The SL light curves and spectra are extracted using the stray light wrappers available in nustar-gen-utils\footnote{\href{https://github.com/NuSTAR/nustar-gen-utils}{GitHub: nustar-gen-utils}} with the region files provided in StrayCats \citep{Grefenstette_2021, Ludlam_2022}. The data were inspected for the presence of Type-I X-ray bursts and increased solar activity but no evidence of either were found, so no further action was required to filter the data.

The 3--20 keV light curve and HID is provided in Figure~\ref{fig:HID}. We provide the light curves and HID of the focused observations for comparison to the behavior exhibited by the SL data. The data from the focused observations were extracted using 2 arcminute regions centered on the source. All light curves were extracted using a time bin size of 128s. The data are color-coded by elapsed time since the beginning of the observation to highlight the temporal correspondence between the HID and the light curve. Based on the characteristics of the light curves and HIDs, the SL data captured the source predominantly in the FB with possible minimal contribution from the SA at the start of the observation. In contrast, the focused data obtained following the SL data are found primarily in the SA with excursions into the FB at the start and end of the observation \citep{Fabio_2024}. We do not analyze the focused data further nor further divide the SL data in order to maximize the S/N.

\begin{figure}[t]
  \centering
  \includegraphics[width=0.48\textwidth, trim=5 0 0 0, clip]{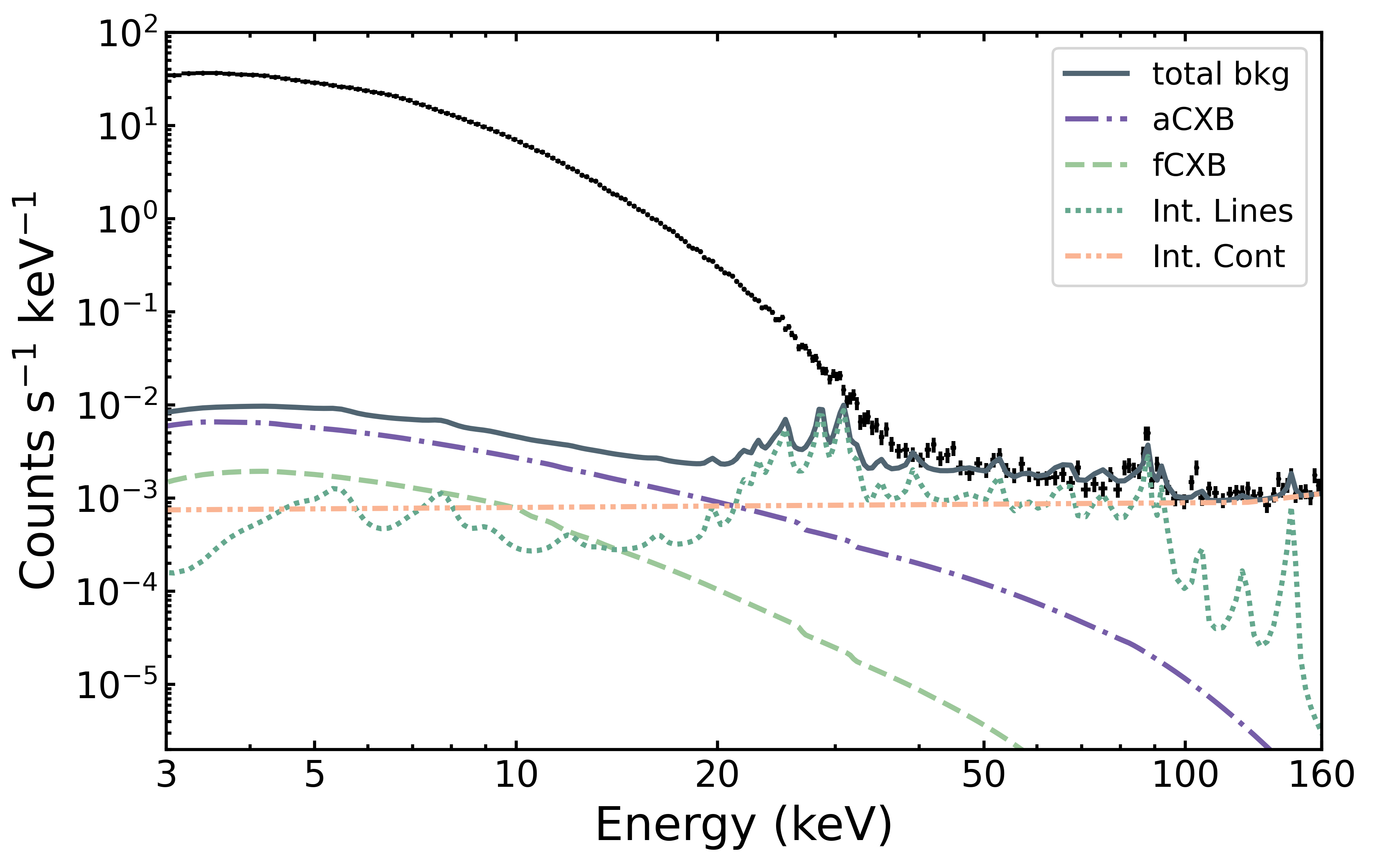}  
  \caption{SL spectrum for FPMA with different background components. The data are shown in black. The dash-dot line is the aCXB, dashed line is the fCXB, dotted line is the internal line emission, and the dot-dot-dashed line is the internal continuum. The overall background contribution is shown as the solid line.}
  \label{fig:background}
\end{figure}

The background estimation for SL observations cannot be extracted from another region on the detectors since there could be faint sources or transmitted light within the background region that could change the shape of the source spectrum when subtracted \citep{madsen_2017, Grefenstette_2021}. 
Therefore, we use {\sc nuskybgd} \citep{nuskybgd} to model the background contributions present in the SL spectrum. {\sc nuskybgd} accounts for both instrumental and astrophysical components including: focused cosmic X-ray background (fCXB), aperture cosmic X-ray (aCXB) background, solar background, and instrumental background. We turned off the solar component since we already determined that the observation does not have increased solar activity. The background analysis is done on the SL region directly. We get an initial estimation on the fCXB and aCXB parameter and then freeze it for the rest of the analysis. Then we fit the internal continuum (Int.\ Cont) in 105 -- 160 keV range and internal lines (Int.\ Lines) in 80 -- 160 keV range where there is no source spectrum. We freeze all the background components during the subsequent source analysis. The relative background contributions to the SL spectrum are shown in Figure~\ref{fig:background}. From comparing the spectrum and background of the intentional SL observation of Sco X-1 in Figure~\ref{fig:background} with the serendipitous SL observations of GX~340+0 (c.f., figures 4 \& 5 in \citealt{songwei_2025}), we can see that the count rate of the intentional SL observation of Sco X-1 is about 2 orders of magnitude higher than the serendipitous SL observations on GX~340+0. This yields a S/N of $> 500$ near 6 keV and S/N~$\sim14$ near 30~keV; thereby enabling robust spectral analysis of the reflection spectrum.

\input{fit}

\section{Spectral analysis \& results}
Spectral analysis was performed using {\sc xspec} version 12.13.0c \citep{arnaud96}. Spectra were binned using the scheme of \cite{kb16} with a minimum number of 30 counts per bin and considered from 3 keV to 37.5 keV, where the background begins to contribute significantly. The FPMA and FPMB spectra were modeled jointly with a constant factor to account for the difference in illuminating area between FPMs. The fits were performed using Cash statistics \citep{cash}. Errors are reported at the 90\% confidence level.
The column density along the line of sight ($N_H$) is modeled with {\sc tbabs} \citep{Wilms_2000} and fixed at $1.5 \times 10^{21}\ \mathrm{cm^{-2}}$ \citep{HI4PI} throughout the analysis.

The continuum was initially modeled with an absorbed double thermal model, yielding a reduced C-statistic of 1308/269, which we rejected as an incomplete description of the source spectrum. Visual inspection showed an additional signature for weak Comptonization, so we explored the use of a phenomenological power-law component. However, it still provided a poor fit, with a reduced C-statistic of $1178/267$. We also tried a more physically motivated Comptonization model with {\sc thcomp} \citep{Zdziarski_2020}, testing two configurations: Model A {\sc tbabs*(bbody+thcomp*diskbb)}, representing Comptonization of disk photons, and Model B {\sc tbabs*(thcomp*bbody+diskbb)}, representing Comptonization of seed photons from the NS surface or BL. They both improved the fit significantly, reducing the C-statistic to $\sim 530/266$. Given this substantial improvement of a Bayesian Information Criterion (BIC) difference $\Delta \rm BIC\sim643$ after using the {\sc thcomp}, we retained the Comptonization component for further analysis. Models A and B give very similar results in both the best fit values and statistic, showing no preferences ($\Delta \rm BIC=3$) on whether the Comptonization originates from the NS surface/BL or the disk. Therefore, we kept both configurations for further analysis.

\begin{figure*}[ht]
  \centering
  \includegraphics[width=0.48\textwidth]{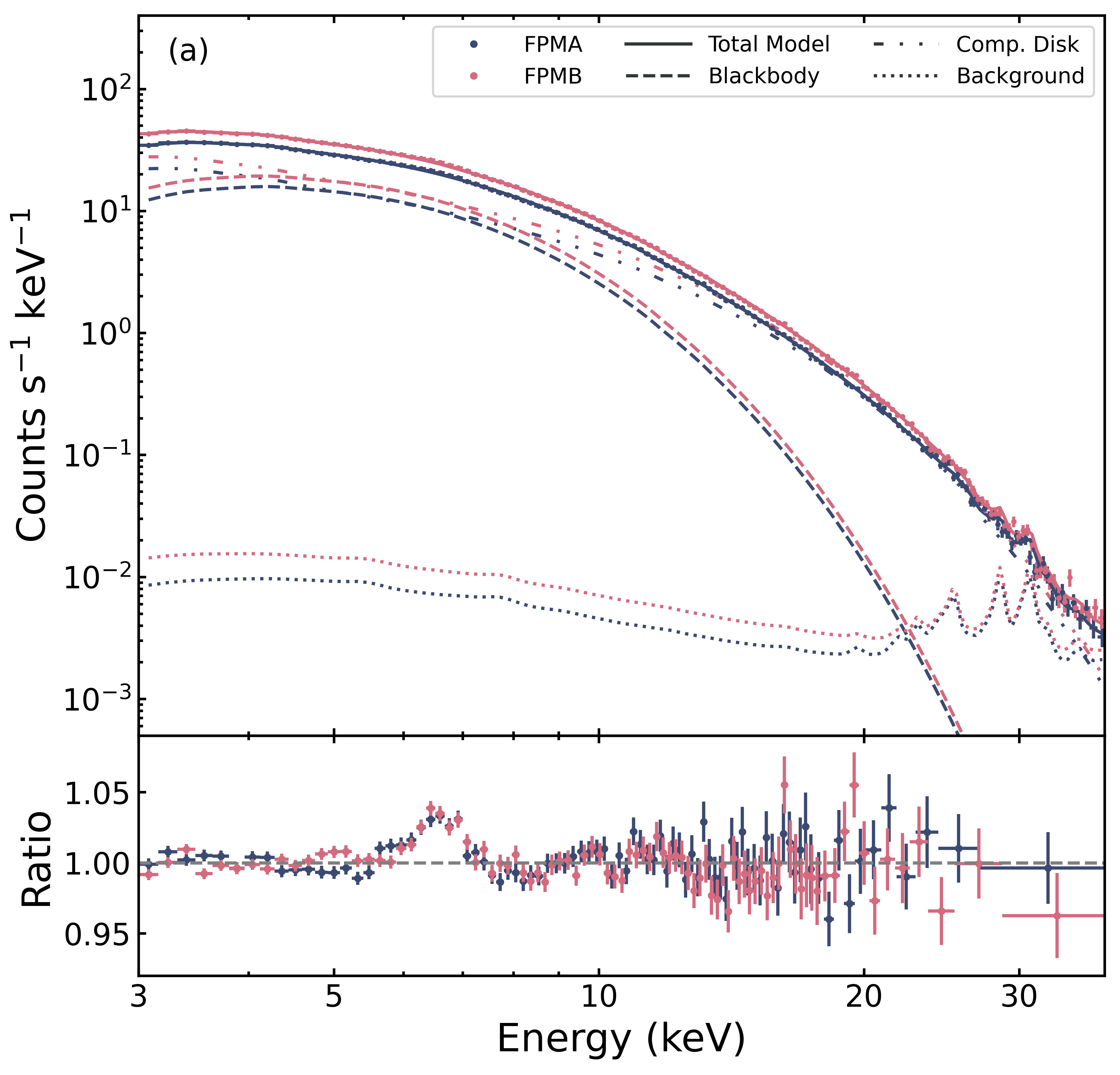}
  \hspace{0.02\textwidth}
  \includegraphics[width=0.48\textwidth]{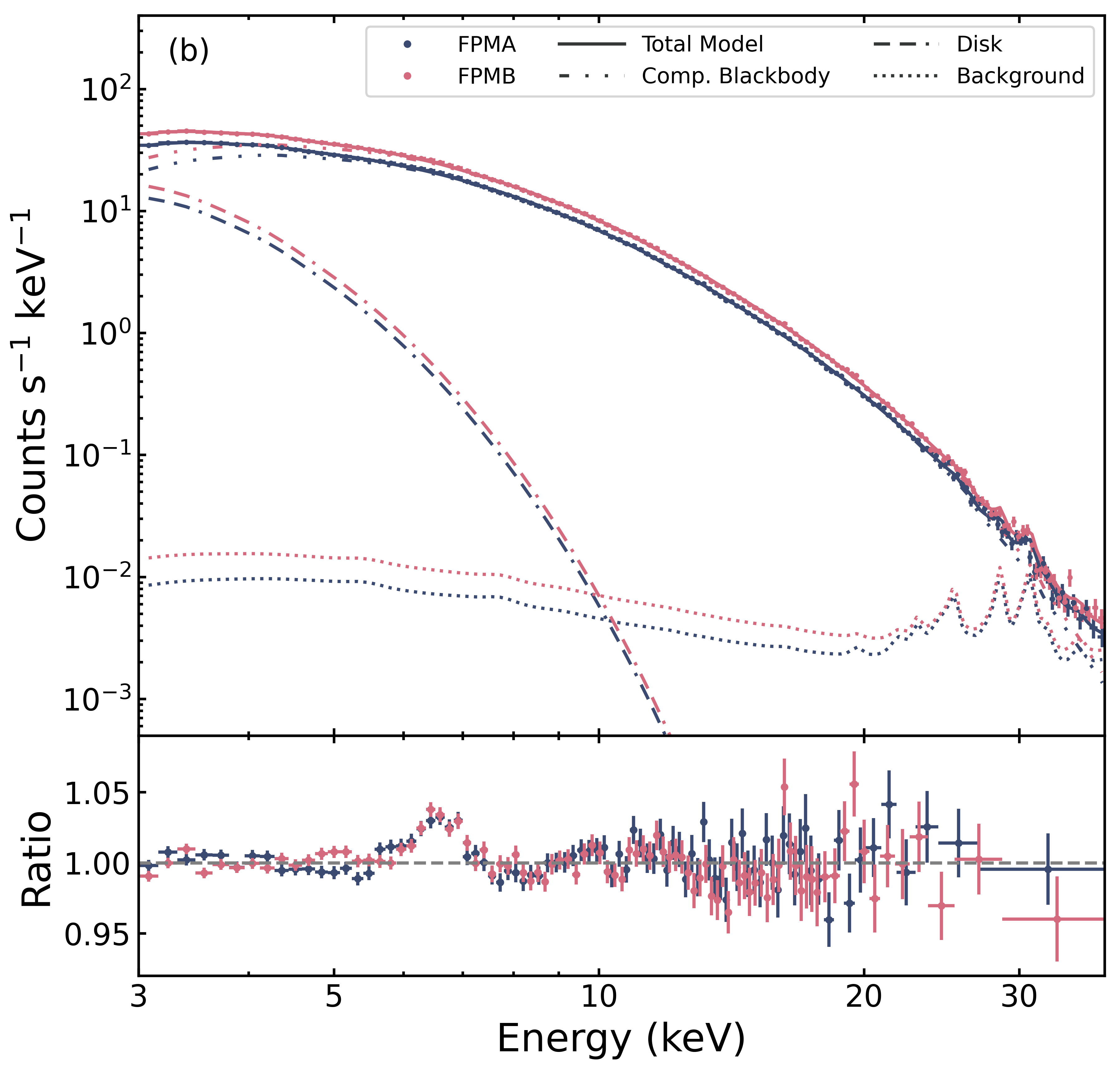}

  \vspace{0.02\textwidth}

  \includegraphics[width=0.48\textwidth]{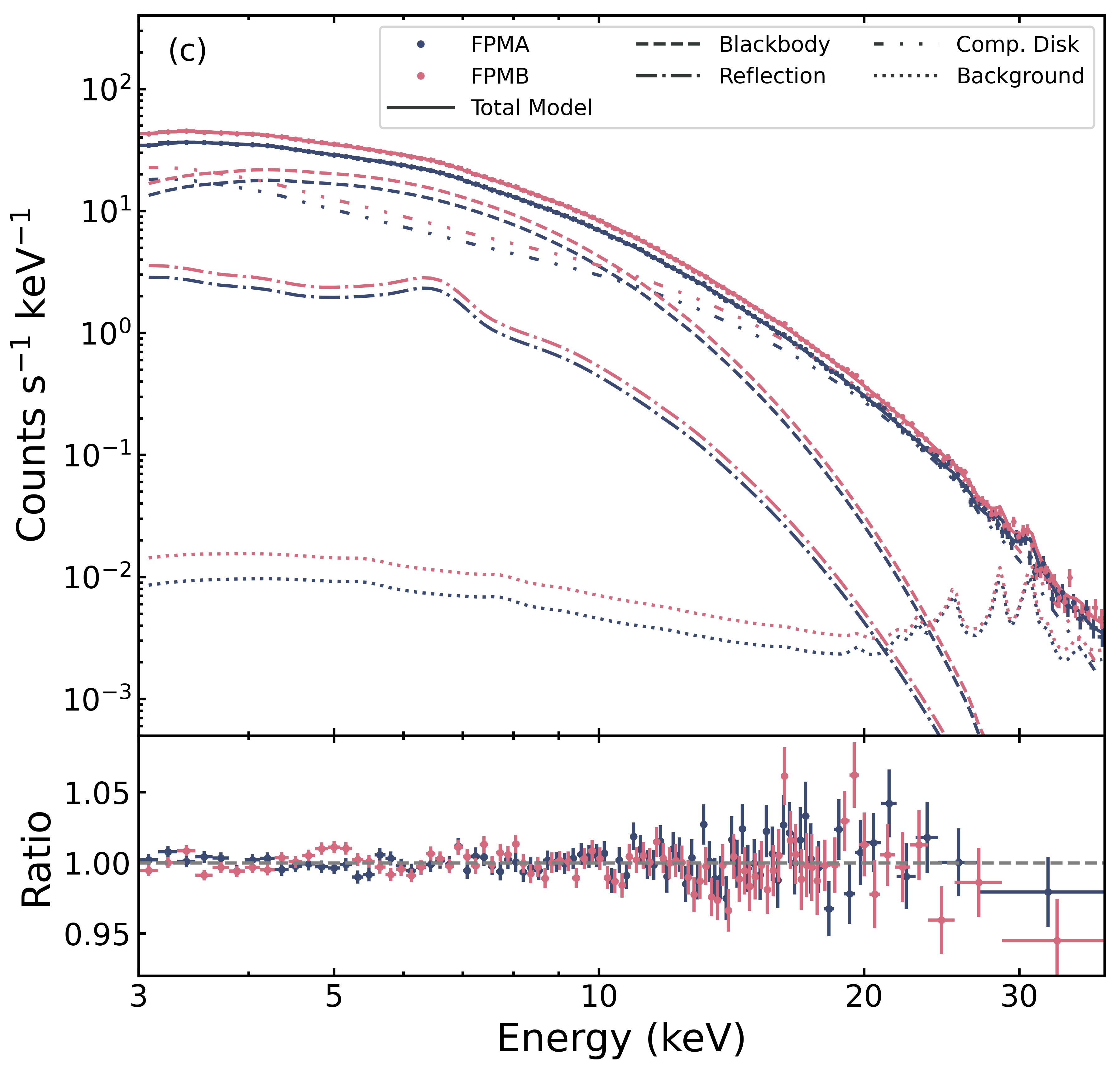}
  \hspace{0.02\textwidth}
  \includegraphics[width=0.48\textwidth]{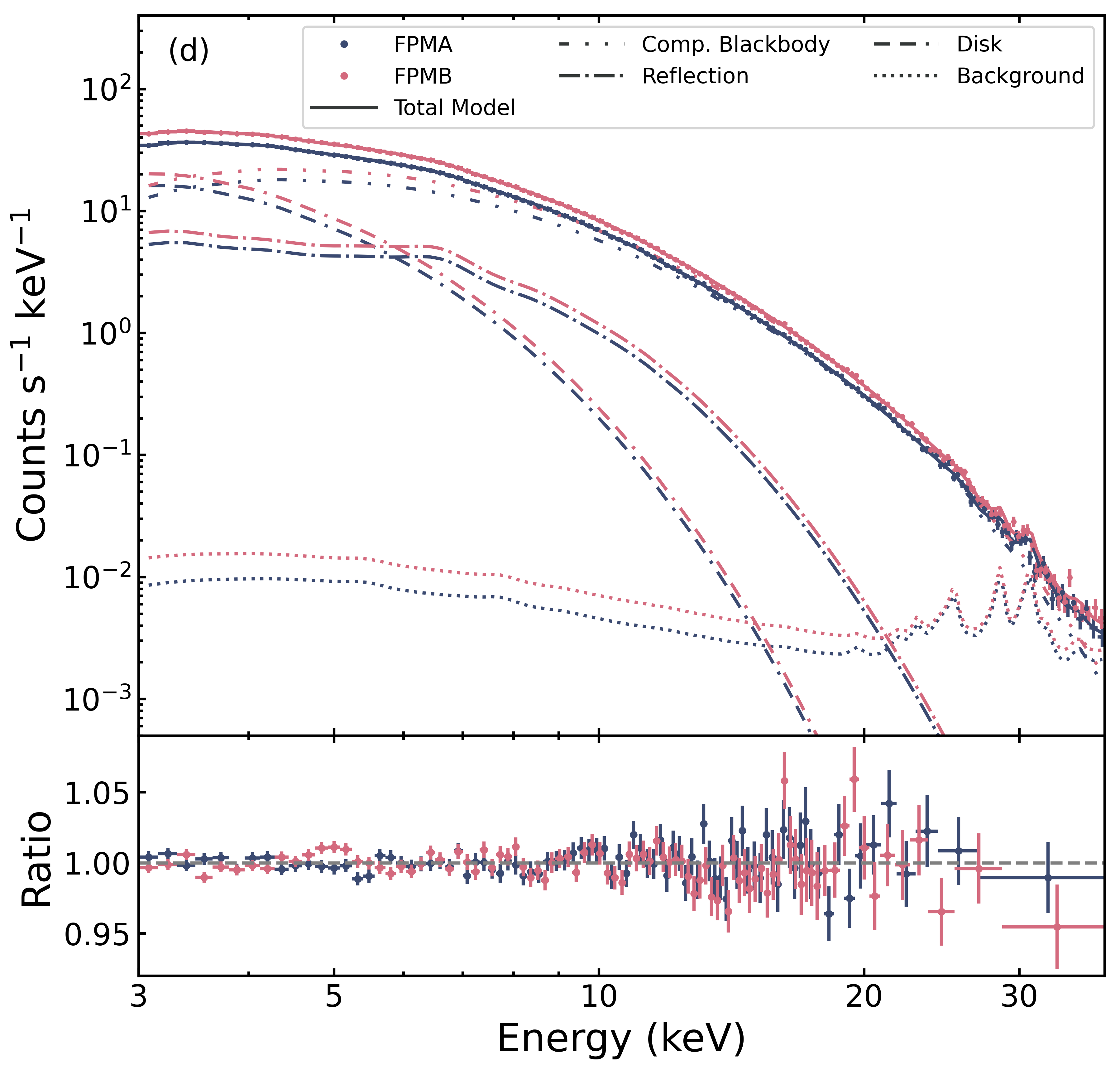}

  \caption{Spectra and ratio of overall model to data for (a) Model A, (b) Model B, (c) Model C, and (d) Model D. In all panels, the spectrum with different components are plotted in the top sub-panel. FPMA is plotted in dark grey and FPMB is plotted in mauve. The total model is plotted in solid line, the total background model is plotted in dotted line, the {\sc bbody} component is plotted in dashed line, the {\sc diskbb} component is plotted in dash-dash-dot line, the convoled component {\sc thcomp} is plotted in dash-dot-dot line, and the {\sc relxillNS} component is plotted in dash-dot line. Data were visually rebinned in the ratio panels for clarity.}
  \label{fig:cont_ratio}
\end{figure*}

We found evidence of an Fe K emission line between 6.4 -- 6.97 keV, which indicates the presence of reprocessed emission in the spectra. 
The addition of a simple Gaussian emission line to Models A and B indicates a line energy of $6.53_{-0.03}^{+0.05}$~keV with line width $\sigma=0.38_{-0.06}^{+0.07}$~keV and $6.53_{-0.06}^{+0.05}$~keV with line width $\sigma=0.39_{-0.06}^{+0.07}$~keV, respectively. The equivalent width of the line is $\sim41$~eV regardless of the continuum model used and detected at $\gtrsim5\sigma$ confidence level. This is similar to the \ion{Fe}{25} line reported in \cite{Mazzola_2021} for the FB of focused \nustar data. While the addition of a Gaussian line improves the overall fit (C-stat/dof~=~296/263 when applied to Model A and C-stat/dof~=~300/263 when applied to Model B), a single emission line is insufficient to fully describe the reprocessed emission in the system nor does it enable determining physical properties of the system.
We applied the reflection model {\sc relxillNS} \citep{Garcia_2022} to account for the reprocessed emission from the accretion disk. Models C and D use the same continuum as Models A and B, respectively, but with the addition of the reflection component. In both cases, the input temperature for the reflection component is set to the blackbody temperature $kT_{bb}$ of the continuum. The $refl_{frac}$ is fixed at $-1$ so the model only accounts for the reflection. We tie the emissivity index $q_2$ to $q_1$ for a single emissivity profile $q$ for illuminating the disk.
The dimensionless spin $a$ is fixed at 0 because it is expected to be smaller than 0.3 for Galactic NS LMXBs \citep{Galloway_2008} and the difference of the inner disk radius between $a=0$ and $a=0.3$ is only about 1~\rg. The outer disk radius is set to $R_{out} = 1000R_{g}$ and the redshift $z$ is fixed at 0 since \source is a Galactic source. The disk density $\log(n_e/ \rm cm^{-3})$ is fixed at the maximum allowed value of 19 in {\sc relxillNS} since the disk electron density in NS systems are expected to be in excess of $10^{20}\ \rm cm^{-3}$ \citep{Frank_2002,Shakura_1973}. 
The parameter values of the best fit results for each model can be found in Table \ref{fit}.

\section{Discussion}
We present the spectral analysis of the first intentional SL observation of Sco X-1 with \nustar, which captured the source in a frequent flaring state. The data from FPMA and FPMB were modeled simultaneously to obtain information about the continuum and reprocessed emission. The continuum was modeled with a combination of thermal components and Comptonized thermal emission. We tested for Comptonized emission arising from the accretion disk or close to the NS. Both configurations provide similar statistical results for the continuum and reflection models.

The blackbody temperatures we found are comparable to the previous results in the FB of focused \nustar observations for \source \citep{Mazzola_2021}. The disk temperature inferred here is slightly elevated in comparison to \cite{Mazzola_2021} and \cite{Fabio_2024}, however, they both used {\sc nthcomp} instead of {\sc thcomp} to account for the Comptonization. Notably, \cite{Mazzola_2021} used a Gaussian line function to model the reflection spectrum which does not account for the reprocessed continuum emission which can skew the relative inferred contributions to the spectrum from the thermal components.

From Figure~\ref{fig:cont_ratio}, it is evident that the thermal Comptonized emission accounted for by the {\sc thcomp} dominates in the hard X-ray regime above 10~keV regardless of the assumed input seed photon distribution. This is consistent with the findings of \citet{Church_2012}. The photon index in {\sc thcomp} varies between $\Gamma\sim$~1.2--2.5 and the electron temperature ranges between $kT_{e}\sim$~2.7--3.0~keV which are consistent with literature values \citep{Titarchuk_2014, Mazzola_2021, Fabio_2024}. The photon index produced by Comptonization is defined as:
\[
\Gamma = \left[ \frac{9}{4} + \frac{1}{\left( \frac{kT_e}{m_e c^2} \right) \, \tau \left(1 + \frac{\tau}{3} \right)} \right]^{1/2} - \frac{1}{2}
\]
thus enabling the estimation of the optical depth \citep{Sunyaev_1980}. The Thomson optical depth can be estimated to be $\gtrsim 10$ for all fits reported in Table~\ref{fit}, which indicates an optically thick Comptonizing region consistent with that reported in \cite{Mazzola_2021} and \cite{Fabio_2024}.

The reflection component indicates that the inclination angle of the system is between $37^{\circ}$--$57^{\circ}$ depending on the choice of Comptonizing medium. This is consistent with the results from radio observations \citep{Fomalont_2001} and the results from polarization measurements \citep{Fabio_2024}. The emissivity index is found to be $q\sim 2$. This is consistent with the reflection modeling results from \cite{Fabio_2024} which used the focused \nustar dataset following the SL observation in conjunction with IXPE, NICER, and HXMT data. The ionization is moderate with a best fit value of $\log(\xi)$ between 1.60--1.92, which is lower than the reported value of $\sim 2.5$ from \cite{Fabio_2024}. The difference could be due to difference in spectral state between the two observations. 
The analysis from \cite{Fabio_2024} is mostly from SA with only short excursions into the FB while our analysis is almost purely in FB. This could lead to different ionization states as the source transits to softer states and with less Comptonized emission. The iron abundance is close to solar values, though less constrained in Model D. Furthermore, the inner disk radius is constrained in Model C to be $R_{in}=2.5_{-0.2}^{+0.7}$ \risco, which indicates a truncated disk, while Model D provides a less precise estimation of the disk between 1.0--8.0~\risco but
consistent with the result from \cite{Fabio_2024} within the 90\% confidence level.  

The thermal components within Model C and D also provide an estimate of the emission radius for the respective component considered. Although there are uncertainties to be considered that can impact the inferred values by a factor of $\sim2$ (see detailed discussion in \citealt{ludlam22}), we provide these for comparison to the results inferred from reflection modeling. We assume a standard correction factor of 1.7 \citep{kubota_2001} and parallax distance measurement of $2.13_{-0.26}^{+0.21}$~kpc when estimating the emission radius of the single-temperature blackbody and disk component. For Model C where we assume Comptonization arises from the accretion disk, the spherical emission radius of the blackbody component is $34.3_{-5.0}^{+4.5}$~km and the inner disk radius from the {\sc diskbb} component is $82_{-30}^{+13}$~km using the values reported in Table~\ref{fit}. Given that the estimate of the inner disk radius may be overestimated by up to a factor of 2.2 \citep{Zimmerman_2005}, these values are roughly consistent with each other. The value inferred from reflection is $31.0_{-2.5}^{+8.7}$~km, assuming a canonical NS mass of 1.4~$M_{\odot}$ and the assumed dimensionless spin parameter of 0, which is consistent with the inferred emission radii of the thermal components. For Model D where the Comptonization is assumed to arise from the NS/BL region, the spherical emission radius of the blackbody component is $39_{-6}^{+5}$~km and the inner disk radius from the {\sc diskbb} component is $60_{-14}^{+40}$~km using the values reported in Table~\ref{fit}. The value inferred from reflection is $14.9_{-2.5}^{+84.3}$~km. Given the large uncertainty on the radius from reflection modeling, the results are generally consistent between all components, but less constraining. We note that a truncated disk in the FB is consistent with literature of the behavior of other Z sources like GX 349+2 \citep{Coughenour_2018} and likely arises from the intense radiation pressure at high mass accretion rate \citep{Mazzola_2021} and unstable nuclear burning on the surface of the NS \citep{Church_2012}.

Due to the extreme brightness of \source, observations have been challenging because it causes pile-up in CCD detectors. For \nustar, the extreme brightness causes high telemetry loads which limits the maximum exposure time to $\sim 20$ ks \citep{Grefenstette_2021}. This makes the SL observation from \nustar a great way for observing extreme bright sources like \source for extended periods of time as it helps reduce the high telemetry loads, and allows longer exposure times so that multiple orbits can be captured to study the parameter variations as a source varies while still maintain relatively high S/N. For example, the SL observation 10902336002 of \source has an elapsed time of $\sim 32$ ks and an exposure time of $\sim 10$ ks. In comparison, the focused observations 30902036002, 30502010002, and 30001040002 have similar elapsed times $(\sim 32-38)$ ks, but yielded only $0.7 - 1$ ks of usable exposures each due to deadtime, which is just a small fraction of what SL achieve.

\section{Conclusion}
We perform a spectral analysis of new observations of \source in the flaring branch. Given the extreme brightness of \source, \nustar intentionally observed the source at an off-axis angle such that SL illuminated the detectors bypassing the focusing optics. We show that the data of the intentional SL observation provides sufficient S/N for high-quality spectral analysis providing $10\times$ more cumulative exposure time than achieved with focused \nustar observations of \source. The spectra were modeled with a combination of a thermal component and thermal Comptonization presenting both scenarios used in literature for Comptonization by the disk or by the NS/boundary layer. Regardless of the continuum description used, a broad Fe emission line was present in the remaining residuals detected at $\gtrsim5\sigma$ confidence level. We apply a relativistic reflection framework to model this component and infer properties about the accreting system.  We find an inclination between $37^{\circ}$--$57^{\circ}$ in agreement with the previous estimate from radio lobe measurements. 
The inner disk radius is likely truncated in the FB, as indicated by continuum and reflection model estimates. This could be due to intense radiation pressure at high mass accretion rate and unstable nuclear burning on the surface of the NS, which is consistent with the behavior of other Sco-like Z sources.
Overall, the results demonstrate that intentional SL observations can recover key spectral parameters that are comparable to focused observations. This supports the broader use of SL data, especially for bright sources like Galactic X-ray binaries during outburst where the telemetry loads are expected to be extremely high.
\\

\noindent {\it Acknowledgments:} This work is supported by NASA under grant No.\ 80NSSC23K0498.

\renewcommand{\thesection}{\Roman{section}}

\bibliography{reference.bib} 
\bibliographystyle{aasjournal}
\end{document}

%% file: fit.tex

\begin{table*}[ht]
\centering
\begin{tabular}{lc|cc|cc}
\hline
Model & Parameters & \multicolumn{2}{c|}{Continuum} & \multicolumn{2}{c}{Reflection}\\
& & Model A & Model B & Model C & Model D \\
\hline
{\sc Constant} & $C_{\rm FPMB}$ & $1.05\pm 0.01$ &$1.05\pm 0.01$ & $1.05\pm{0.01}$ & ${1.05}\pm 0.01$\\
{\sc thcomp}
&$\Gamma$ &$1.62^{+0.06}_{-0.09}$&$2.5^{+0.2}_{-0.4}$&${1.2}^{+0.1}_{-0.2}$&${1.9}^{+0.4}_{-0.3}$\\
&$kT_{e}$ (keV) &$2.74\pm0.04$&$3.0^{+0.06}_{-0.07}$&${2.70}^{+0.06}_{-0.05}$&${2.86}^{+0.12}_{-0.08}$\\
&$cov_{frac}$&$0.59^{+0.04}_{-0.09}$&$0.76^{+0.08}_{-0.20}$&${0.27}^{+0.02}_{-0.08}$ & ${0.48}^{+0.20}_{-0.09}$ \\
{\sc diskbb}
 & $kT_{in}$ (keV) & $0.59^{+0.06}_{-0.03}$ & $0.63^{+0.06}_{-0.02}$ & ${1.01}^{+0.08}_{-0.01}$ & ${1.06}^{+0.01}_{-0.09}$ \\
 & $\rm norm\ (10^{4})$ & $20.3^{+11.2}_{-8.8}$ & $8.1^{+1.5}_{-2.0}$ & ${1.2}^{+0.04}_{-0.35}$ & ${0.7}^{+0.4}_{-0.1}$\\
 {\sc bbody} 
 & $kT_{bb}$ (keV) & $1.47^{+0.02}_{-0.03}$ & $1.33\pm0.03$ &${1.58}^{+0.05}_{-0.02}$&${1.55}^{+0.02}_{-0.08}$\\
 & norm & $2.3^{+0.2}_{-0.3}$ & $3.93^{+0.03}_{-0.02}$ & ${2.5}\pm0.2$ & ${3.0}\pm0.2$ \\
{\sc relxillNS}
 &$q$ & ... & ... & ${2.2}\pm0.7$ & ${2.1}^{+0.4}_{-0.9}$ \\
 &$i$ ($^\circ$) & ... & ... & ${48}^{+6}_{-8}$ & ${43}^{+14}_{-6}$ \\
 & $R_{in}$ (\risco) & ... & ... & ${2.5}^{+0.7}_{-0.2}$ & ${1.2}^{+6.8}_{-0.2}$ \\
 & $\log(\xi)$ & ... & ... & ${1.8}\pm0.2$ &${1.87}^{+0.05}_{-0.21}$\\
 & $A_{Fe}$ & ... & ... & $0.7\pm0.1$ & $<0.9$ \\
 & $\rm norm\ (10^{-2})$ & ... & ... & ${8}\pm3$ & ${11}^{+5}_{-3}$\\
 \hline
\multirow{1}{*}{Statistic} 
 & C-stat/dof & 529/266 & 526/266 & 280/258 & 281/258\\
\hline
\end{tabular}
\hspace*{0.001\textwidth}
\parbox{.9\textwidth}{
\caption{Spectral modeling results for Sco X-1. Errors are reported at 90\% confidence level. The column density $N_H$ is fixed at $1.5\times10^{21}\ \mathrm{cm^{-2}}$. Model A assumes a Comptonized disk, whereas Model B assumes a Comptonized blackbody in the continuum. Model C and D have a Comptonized disk and blackbody component, respectively, with reflection assuming to arise from illumination of the blackbody component (i.e., $kT_{bb}$ of \relxillns is tied to the temperature of the {\sc bbody} component). The emissivity index $q_2$ is tied to $q_1$. 
The outer disk radius $R_{out}$ is fixed at 1000 \rg, the spin is fixed at 0, and the reflection fraction is set to $-1$ to return the reflection spectrum only in the \relxillns model.}
\label{fit}
}
\end{table*}